\documentclass[conference,10pt]{IEEEtran}
\IEEEoverridecommandlockouts
\usepackage{lipsum} 

\usepackage{graphicx}
\usepackage{cite}
\usepackage{amsfonts} 
\usepackage{amsmath}  
\usepackage{bm}
\usepackage{psfrag}
\usepackage{amsbsy}
\usepackage{amssymb}
\usepackage{mathrsfs}
\usepackage{stfloats}
\usepackage{dsfont}
\usepackage{setspace}
\usepackage{array}
\usepackage{bbm}
\usepackage{mathabx}
\usepackage{algorithmic}
\usepackage{algorithm}
\usepackage{glossaries}
\usepackage{color}
\usepackage{amsthm}
\usepackage{lettrine}

\newtheoremstyle{mystyle}
{}                
{}                
{\itshape}        
{}                
{\bfseries}       
{:}               
{.5em}            
{}                

\theoremstyle{mystyle}
\newtheorem{proposition}{Proposition}

\newcommand{\ds}{\displaystyle}

\usepackage{caption}
\DeclareMathOperator{\tr}{tr}

\def\BibTeX{{\rm B\kern-.05em{\sc i\kern-.025em b}\kern-.08em
		T\kern-.1667em\lower.7ex\hbox{E}\kern-.125emX}}
	
\usepackage[left=0.625in, right=0.625in, top=0.75in, bottom=1in]{geometry}

% ...

\usepackage{enumitem}
\setlist[itemize]{label=$\bullet$}

\usepackage{hyperref}

\hypersetup{
	colorlinks=true,       
	linkcolor=blue,       
	citecolor=green,      
	filecolor=magenta,    
	urlcolor=cyan         
}

\usepackage{caption}
\usepackage{url}
\makeatletter
\def\url@remove@extrahref#1{%
	\xdef\@savedhref{#1}%
	\expandafter\expandafter\expandafter\strip@url
	\expandafter\@savedhref
	\expandafter\@gobble
}
\makeatother

\makeglossaries

\newacronym{mac}{MAC}{multiple-access channel}
\newacronym{bc}{BC}{broadcast channel}
\newacronym{mimo}{MIMO}{multiple-input multiple-output}
\newacronym{siso}{SISO}{single-input single-output}
\newacronym{sc}{SC}{single-carrier}
\newacronym{mc}{MC}{multi-carrier}
\newacronym{ofdma}{OFDMA}{orthogonal frequency division multiple access}
\newacronym{af}{AF}{amplify-and-forward}
\newacronym{df}{DF}{decode-and-forward}
\newacronym{cf}{CF}{compress-and-forward}
\newacronym{mwrc}{MWRC}{multi-way relay channel}
\newacronym{pde}{PDE}{partial data exchange}
\newacronym{fde}{FDE}{full data exchange}
\newacronym{iid}{i.i.d.\@}{independent and identically distributed}
\newacronym{awgn}{AWGN}{additive white Gaussian noise}
\newacronym{awg}{AWG}{additive white Gaussian}
\newacronym{sic}{SIC}{successive interference cancellation}
\newacronym{snr}{SNR}{signal-to-noise ratio}
\newacronym{sinr}{SINR}{signal-to-interference-plus-noise ratio}
\newacronym{ber}{BER}{bit error rate}
\newacronym{zf}{ZF}{zero-forcing}
\newacronym{mmse}{MMSE}{minimum mean square error}
\newacronym{sud}{SUD}{single user decoding}
\newacronym{dof}{DoF}{degrees of freedom}
\newacronym{gdof}{GDoF}{generalized degrees of freedom}
\newacronym{nnc}{NNC}{noisy network coding}
\newacronym{dmn}{DMN}{discrete memoryless network}
\newacronym{csi}{CSI}{channel state information}
\newacronym{ee}{EE}{energy efficiency}
\newacronym{ian}{IAN}{treating interference as noise}
\newacronym{snd}{SND}{simultaneous non-unique decoding}
\newacronym{brd}{BRD}{best response dynamics}
\newacronym{br}{BR}{best response}
\newacronym{ne}{NE}{Nash equilibrium}
\newacronym{lhs}{LHS}{left-hand side}
\newacronym{rhs}{RHS}{right-hand side}
\newacronym{gee}{GEE}{global energy efficiency}
\newacronym{wsee}{WSEE}{weighted sum energy efficiency}
\newacronym{wpee}{WPEE}{weighted product energy efficiency}
\newacronym{wmee}{WMEE}{weighted minimum energy efficiency}
\newacronym{kkt}{KKT}{Karush-Kuhn-Tucker}
\newacronym{pc}{PC}{pseudo-concave}
\newacronym{qc}{QC}{quasi-concave}
\newacronym{ql}{QL}{quasi-linear}
\newacronym{pl}{PL}{pseudo-linear}
\newacronym{spc}{SPC}{strictly pseudo-concave}
\newacronym{sqc}{SQC}{strictly quasi-concave}
\newacronym{lfp}{LFP}{linear fractional problem}
\newacronym{clfp}{CLFP}{concave-linear fractional problem}
\newacronym{ccfp}{CCFP}{concave-convex fractional problem}
\newacronym{mmfp}{MMFP}{max-min fractional problem}
\newacronym{sorp}{SoRP}{sum-of-ratios problem}
\newacronym{porp}{PoRP}{product-of-ratios problem}
\newacronym{srp}{SRP}{single-ratio problem}
\newacronym{brb}{BRB}{branch-reduce-and-bound}
\newacronym{qos}{QoS}{quality-of-service}
\newacronym{comp}{CoMP}{cooperative multi-point}
\newacronym{ue}{UE}{user equipment}
\newacronym{bs}{BS}{base station}
\newacronym{mrc}{MRC}{maximum ratio combining}
\newacronym{d2d}{D2D}{device-to-device}
\newacronym{lmmse}{LMMSE}{linear minimum mean square error}
\newacronym{ris}{RIS}{reconfigurable intelligent surface}
\newacronym{svd}{SVD}{singular values decomposition}

\title{Secrecy Energy Efficiency Maximization in RIS-Aided Wireless Networks with Statistical CSI}

\author{
	Robert Kuku Fotock, {\em Student Member, IEEE}, Agbotiname Lucky Imoize, {\em Senior Member, IEEE},  Alessio Zappone,\\ {\em Senior Member, IEEE}, Marco Di Renzo, {\em Fellow, IEEE}, Roberto Garello, {\em Senior Member, IEEE}
	\thanks{R. K. Fotock and A. Zappone are with the University of Cassino and Southern Lazio and with CNIT, Italy (\{robertkuku.fotock, alessio.zappone\}@unicas.it). A. L. Imoize is with CNIT and with Politecnico di Torino, Italy (agbotiname.imoize@polito.it). M. Di Renzo is with Universit\'e Paris-Saclay, CNRS, CentraleSup\'elec, Laboratoire des Signaux et Syst\`emes, France (marco.di-renzo@universite-paris-saclay.fr). R. Garello is with Politecnico di Torino, Italy (roberto.garello@polito.it).
		The work of R. K. Fotock and A. L. Imoize was supported by the European Commission through the H2020-MSCA-ITN-METAWIRELESS project, grant agreement 956256, and the HE-MSCA-DN-INTEGRATE project, grant agreement number 101072924, respectively. The work of A. Zappone was supported by the SPARKS Project with CUP D43C22003080001, within the framework of the (NRRP) of NextGenerationEU, partnership on “Telecommunications of the Future” (PE00000001 - program “RESTART”). The work of M. Di Renzo was supported in part by the European Commission through the H2020 ARIADNE project, grant agreement 871464, and the H2020 RISE-6G project, grant agreement 101017011.
}}
\begin{document}
	
	\sloppy  
	
	\maketitle
	
	\begin{abstract}
		
This work studies the problem of secrecy energy efficiency maximization in multi-user wireless networks aided by reconfigurable intelligent surfaces, in which an eavesdropper overhears the uplink communication. A provably convergent optimization algorithm is proposed which optimizes the user's transmit power, metasurface reflection coefficients, and base station receive filters. The complexity of the proposed method is analyzed and numerical results are provided to show the performance of the proposed optimization method.

	\end{abstract}
	
	\vspace{1em}
	
	
	\section{Introduction}
\lettrine[nindent=0.1em,lines=2]{R}{ECONFIGURABLE} intelligent surfaces (RISs) have been proposed as a promising technology for 6G networks, as they provide satisfactory rate performance with lower energy consumption than traditional antenna arrays \cite{DiRenzo2020, SmartWireless, RuiZhang_COMMAG, Huang2019}. This can unlock unprecedented levels of energy efficiency (EE), a key performance indicator of 6G  networks~\cite{Lopezperez2021survey}. Moreover, aiming at improving their rate gains, recently it has been proposed to deploy dedicated analog amplifiers on the RIS, which led the research community to investigate the use of active RISs~\cite{Long2021}. However, this poses a fundamental trade-off in terms of EE, since equipping the RIS with additional hardware also causes larger energy consumption. Thus, the EE advantages of active RIS are not clear. In~\cite{Fotock2023}, active and nearly passive RISs have been compared in terms of EE, showing that active RISs do not always provide better EE. 

This work continues the investigation started in \cite{Fotock2023}, also focusing on another major requirement of future wireless networks, ensuring secret communications. Several contributions have reported RISs in conjunction with physical layer security techniques. However, most contributions consider the maximization of the system secrecy rate, without addressing the energy efficiency of the network. A non-orthogonal multiple access (NOMA) network that employs a simultaneous transmission and reception (STAR) RIS is considered in~\cite{Li2023}, including the characterization of the system secrecy outage probability. A NOMA-based network aided by a STAR-RIS is also considered in~\cite{Pei2023}, and closed-form approximations of the secrecy outage probability are derived. In~\cite{Zhang2022}, a STAR-aided NOMA-based network is considered and the worst-case secrecy capacity of the system is maximized. In~\cite{Wei2022}, the ergodic secrecy capacity of a RIS-aided wireless network is analyzed and approximated in closed form, considering flying eavesdroppers. In~\cite{Zhao2023}, the secrecy maximization rate of a RIS-aided network powered by wireless power transfer is optimized. In~\cite{Hoang2023}, the secrecy rate of a RIS-aided network with space-ground communications is optimized. In~\cite{Yizhi2023}, the sum secrecy rate of the uplink of a multi-user RIS-aided wireless network is addressed. The SEE in RIS-aided networks has been previously considered in fewer works, such as \cite{Yichi2023} and \cite{Yang2023}. Specifically, in~\cite{Yichi2023}, a deep reinforcement learning method is employed to optimize the SEE of a RIS-aided network, while in~\cite{Yang2023}, a combination of alternating maximization and sequential programming is employed to maximize the SEE of a multi-user network. 

This work aims at analyzing and optimizing the secrecy energy efficiency (SEE)~\cite{ZapTSP13}, in the uplink of a multi-user network in the presence of an eavesdropper. Assuming that only statistical channel state information (CSI) is available for the eavesdropper's channel, a provably convergent optimization algorithm is developed which optimizes the mobile users' transmit powers, the RIS reflection coefficients and the legitimate base station receive filters. The developed model and optimization algorithm are suited to both nearly passive and active RISs, thus allowing us to evaluate the energy efficiency of both approaches. Moreover, the proposed model is general enough to encompass the recently proposed use of RIS with global reflection capabilities. This new kind of RIS generalizes the use of traditional RISs with local reflection capabilities since the constraint on the reflected power is not applied to each reflecting element individually, but rather to the complete surface~\cite{Renzo2022}.

%
	
\section{System Model and Problem Formulation}\label{Sec:SysModel}
\noindent Let us consider a network consisting of $K$ single-antenna mobile transmitters, labeled as Alice, which communicate with their base station, labeled as Bob, equipped with $N_{B}$ antennas, through a reconfigurable intelligent surface (RIS), equipped with $N$ reflecting elements. In the same area, a single-antenna eavesdropper is present, which is labeled as Eve. Let us denote by $p_{k}$ the $k$-th user's transmit power, by $\boldsymbol{\gamma} = \left(\gamma_{1},\ldots,\gamma_{N}\right)$, the $N \times 1$ vector containing the RIS reflection coefficients, by $\boldsymbol{h}_{k}$ the $N \times 1 $ channel between the $k$-th user and the RIS and by $\boldsymbol{G}_{B}$ and $\boldsymbol{g}_{E}$, the $N_{B} \times N$ and $N \times 1$ channel from the RIS to Bob and Eve, respectively. Thus, the SINRs of user $k$ at the intended and eavesdropping receiver, upon linear reception by $\boldsymbol{c}_{k,B}$ and $\boldsymbol{c}_{k,E}$, are written as 
\begin{align}\label{Eq:SINR_B}
&\hspace{-0.15cm}\text{SINR}_{k,B}=\frac{p_{k}\left|\boldsymbol{c}_{k,B}^{H}\boldsymbol{A}_{k,B}\boldsymbol{\gamma}\right|^{2}}{\boldsymbol{c}_{k,B}^{H}\boldsymbol{W}_{B}\boldsymbol{c}_{k,B}+\sum_{m\neq k}p_{m}\left|\boldsymbol{c}_{k,B}^{H}\boldsymbol{A}_{m,B}\boldsymbol{\gamma}\right|^{2}}\\
\label{Eq:SINR_E}
&\hspace{-0.15cm}\text{SINR}_{k,E}\!=\!\frac{p_{k}\left|\boldsymbol{g}_{E}^{H}\boldsymbol{H}_{k}\boldsymbol{\gamma}\right|^{2}}{\sigma_{E}^{2}\!+\!\sigma_{RIS}^{2}\boldsymbol{g}_{E}^{H}\boldsymbol{\Gamma}\boldsymbol{\Gamma}^{H}\boldsymbol{g}_{E}\!+\!\sum_{m\neq k}p_{m}\left|\boldsymbol{g}_{E}^{H}\boldsymbol{H}_{m}\boldsymbol{\gamma}\right|^{2}}
\end{align}
wherein $\boldsymbol{A}_{k,B}=\boldsymbol{G}_{B}\boldsymbol{H}_{k}$ and $\boldsymbol{W}_{B} = \sigma_{B}^{2}\boldsymbol{I}_{N_{B}} + \sigma_{RIS}^{2}\boldsymbol{G_{B}}\boldsymbol{\Gamma}\boldsymbol{\Gamma}^{H}\boldsymbol{G}_{B}^{H}$ is the noise covariance matrix at Bob, with $\boldsymbol{\Gamma} = \text{diag}\left(\boldsymbol{\gamma}\right)$, $\sigma^{2}_{\text{RIS}}$ the noise variance at the RIS, while $\sigma^{2}_{\text{B}}$ and $\sigma^{2}_{\text{E}}$ are the noise variance at Bob and Eve, respectively.
In this context, the goal is to optimize the system SEE, defined as the ratio of the system secrecy rate over the total power consumed in the network. As for the system secrecy rate, it is defined as $R_{s}=\max\{0,\sum_{k=1}^{K}\log_{2}(1+\text{SINR}_{k,B})-\log_{2}(1+\text{SINR}_{k,E})\}$. As for the power consumption, it is obtained by summing the radiofrequency power consumed by the RIS and the users’ transmit power, plus the static power consumption of the whole legitimate system, which yields $P_{tot} = P_{RIS}  + \sum_{k=1}^{K}\mu_{k}p_{k} + P_{c}$, wherein $\mu_{k}$ denotes the inverse efficiency of the transmit amplifier associated with the $k$-th transmitter, and $P_{c} = NP_{c,n} + P^{\text{RIS}}_{0} + P_{0}$, with $P_{c,n} $ the static power consumption of the $n$-th RIS element, $P^{\text{RIS}}_{0}$ is the rest of the static power consumed by the RIS and $P_{0}$ encompasses all other sources of power consumption in the legitimate system.  As for $P_{RIS}$, it is given by the difference between the incident power on the RIS $P_{in}$ and the power that departs from the RIS, $P_{out}$, which after some elaborations~\cite{Robert2023}, can be computed as;
\begin{align}\label{Eqn:P_RF}
	P_{out} - P_{in} &= \tr\left(\textstyle \sum_{k=1}^{K}p_{k}\boldsymbol{\Gamma}\boldsymbol{h}_{k}\boldsymbol{h}^{H}_{k}\boldsymbol{\Gamma}^{H} +  \sigma^{2}_{\text{RIS}}\boldsymbol{\Gamma}\boldsymbol{\Gamma}^{H}\right)\\
	&- \textstyle \sum_{k=1}^{K}p_{k} \left\Vert \boldsymbol{h}_{k}\right\Vert^{2} - \sigma^{2}_{\text{RIS}}N = \tr\left(\left(\boldsymbol{\gamma}\boldsymbol{\gamma}^{H}-\boldsymbol{I}_{N}\right)\boldsymbol{R}\right)\notag
\end{align}
wherein $\boldsymbol{R}=\textstyle\sum_{k=1}^{K} p_{k}\boldsymbol{H}_{k}^{H}\boldsymbol{H}_{k}+\sigma_{\text{RIS}}^{2}\boldsymbol{I}_{N}$.  
The above power consumption model has been developed for an active RIS, but it is general enough to admit the case of a nearly passive RIS as a special case. Specifically, in the case of a nearly passive RIS, $P_{out} \leq P_{in}$ and thus the difference $P_{out} - P_{in}$ does not appear in the power consumption $P_{tot}$. Moreover, if a nearly passive RIS is employed, the terms $P_{c,n}$ and $ P^{\text{RIS}}_{0}$ will have a lower numerical value compared to the case of an active RIS since simpler hardware is employed in nearly-passive RIS, i.e. no analog amplifiers are used. Similarly, the constraints that should be enforced on the RIS vector $\boldsymbol{\gamma}$ depend on whether the RIS is active or nearly passive. Specifically, if the RIS is active, then $0 \leq P_{out} - P_{in} \leq P_{R,\text{max}}$ with $P_{R,\text{max}}$ the maximum radio-frequency power that the RIS amplifier can provide. Instead, for nearly-passive RISs, the constraint $P_{out} \leq P_{in}$ applies, which is a special case of the constraint in the active case, obtained by relaxing the first inequality and setting $\sigma^{2}_{\text{RIS}} = 0\, \text{and} \, P_{R,\text{max}} = 0$ in the second inequality. In the following, the focus will be on the more general active RIS scenario. In the sequel, we assume perfect channel state information (CSI) for all channels except the channel $\boldsymbol{g}_{E}$ from the RIS to the eavesdropper, motivated by the fact that the eavesdropper might be a hidden node. Specifically, as for the channel $\boldsymbol{g_{E}}$, we follow a mean feedback model, in which we assume that the legitimate system has only access to the mean value of $\boldsymbol{g}_{E}$, denoted by $\boldsymbol{\widehat{g}}_{E}$, while the true channel is given by $\boldsymbol{g}_{E}=\boldsymbol{\widehat{g}}_{E}+\boldsymbol{\delta}$, with  $\boldsymbol{\delta}\sim {\cal CN}(\boldsymbol{0},\sigma_{g}^{2}\boldsymbol{I}_{N_{E}})$. As a consequence, the legitimate system can not directly maximize the SEE, and an average version of the secrecy rate at the numerator of the SEE should be considered. Namely, the considered problem can be formulated as 
\begin{subequations}\label{Prob:SEE_ARIS}
	\begin{align}
		&\ds\max_{\boldsymbol{\gamma},\boldsymbol{p},\boldsymbol{C}_{B}}\;\frac{\sum_{k=1}^{K}\log_{2}(1+\text{SINR}_{k,B})-\mathbb{E}_{\boldsymbol{\delta}}\left[\log_{2}(1+\text{SINR}_{k,E})\right]}{P_{tot}}\label{Prob:aSEE_ARIS}\\
		&\;\text{s.t.}\;\tr\left(\boldsymbol{R}\right)\leq \tr(\boldsymbol{R}\boldsymbol{\gamma}\boldsymbol{\gamma}^{H})\leq P_{R,max}+\tr\left(\boldsymbol{R}\right)\label{Prob:bSEE_ARIS}\\
		&\;\quad\;\;0\leq p_{k}\leq P_{max,k}\;\forall\;k=1,\ldots,K\;,\label{Prob:cSEE_ARIS}
	\end{align}
\end{subequations}
with $\boldsymbol{C}_{B} = \left[\boldsymbol{c}_{1,B},\ldots,\boldsymbol{c}_{K,B} \right]$, and we have dropped the positive operator $^{+}$, since we assume that the maximum of the secrecy rate is positive\footnote{The case in which the maximum of the secrecy rate is not positive is of no practical interest since it would imply that even with the best radio resource allocation, no secret communication is possible.}.  We also observe that Problem~\eqref{Prob:bSEE_ARIS} is always feasible, since setting $\left|\gamma_{n}\right|=1$ for all $n$ fulfills all constraints.

\section{Proposed optimization method}
The considered problem \eqref{Prob:SEE_ARIS} is challenging due to the non-convexity of the objective function and constraints, and also due to the presence of the statistical expectation in the numerator of the objective. Let us first deal with the statistical expectation. 
Unfortunately, a closed-form expression of the term $\mathbb{E}_{\boldsymbol{\delta}}\left[\log_{2}(1+\text{SINR}_{k,E})\right]$ is not available. Thus, in order to simplify the problem, we resort to the popular approach of approximating the objective by taking the statistical expectation inside the logarithm, which yields
\begin{align}
&\mathbb{E}_{\boldsymbol{\delta}}\left[\log_{2}(1+\text{SINR}_{k,E})\right]\label{Eq:ApproxErgRate}\\
&=\mathbb{E}_{\boldsymbol{\delta}}\left[\log_{2}\left(\sigma_{E}^{2}\!+\!\sigma_{RIS}^{2}\boldsymbol{g}_{E}^{H}\boldsymbol{\Gamma}\boldsymbol{\Gamma}^{H}\boldsymbol{g}_{E}\!+\!\sum_{m=1}^{K}p_{m}\left|\boldsymbol{g}_{E}^{H}\boldsymbol{H}_{m}\boldsymbol{\gamma}\right|^{2}\right)\right]\notag\\
&-\mathbb{E}_{\boldsymbol{\delta}}\!\left[\!\log_{2}\left(\sigma_{E}^{2}\!+\!\sigma_{RIS}^{2}\boldsymbol{g}_{E}^{H}\boldsymbol{\Gamma}\boldsymbol{\Gamma}^{H}\boldsymbol{g}_{E}\!+\!\sum_{m\neq k}p_{m}\left|\boldsymbol{g}_{E}^{H}\boldsymbol{H}_{m}\boldsymbol{\gamma}\right|^{2}\!\right)\!\right]\notag\\
&\approx \log_{2}\left(\sigma_{E}^{2}\!+\!\mathbb{E}_{\boldsymbol{\delta}}\!\left[\!\sigma_{RIS}^{2}\boldsymbol{g}_{E}^{H}\boldsymbol{\Gamma}\boldsymbol{\Gamma}^{H}\boldsymbol{g}_{E}\!+\!\sum_{m=1}^{K}p_{m}\left|\boldsymbol{g}_{E}^{H}\boldsymbol{H}_{m}\boldsymbol{\gamma}\right|^{2}\!\right]\!\right)\notag\\
&-\log_{2}\left(\sigma_{E}^{2}\!+\!\mathbb{E}_{\boldsymbol{\delta}}\!\left[\!\sigma_{RIS}^{2}\boldsymbol{g}_{E}^{H}\boldsymbol{\Gamma}\boldsymbol{\Gamma}^{H}\boldsymbol{g}_{E}\!+\!\sum_{m\neq k}p_{m}\left|\boldsymbol{g}_{E}^{H}\boldsymbol{H}_{m}\boldsymbol{\gamma}\right|^{2}\!\right]\!\right)\notag
\end{align}
Since $\boldsymbol{g}_{E}\!=\!\boldsymbol{\widehat{g}}_{E}\!+\!\boldsymbol{\delta}$, elaborating, we obtain, for all $m=1,\ldots,K$
\begin{align}
&\mathbb{E}_{\boldsymbol{\Delta}}\left[\left|\boldsymbol{g}_{E}^{H}\boldsymbol{H}_{m}\boldsymbol{\gamma}\right|^{2}\right]=\boldsymbol{\widehat{g}}_{E}^{H}\boldsymbol{H}_{m}\boldsymbol{\gamma}\boldsymbol{\gamma}^{H}\boldsymbol{H}_{m}^{H}\boldsymbol{\widehat{g}}_{E}+\sigma_{g}^{2}\|\boldsymbol{H}_{m}\boldsymbol{\gamma}\|^{2}\notag\\
&=\boldsymbol{\gamma}^{H}\boldsymbol{H}_{m}^{H}\left(\boldsymbol{\widehat{g}}_{E}\boldsymbol{\widehat{g}}_{E}^{H}+\sigma_{g}^{2}\boldsymbol{I}_{N}\right)\boldsymbol{H}_{m}\boldsymbol{\gamma}=\|\boldsymbol{R}_{E}^{1/2}\boldsymbol{H}_{m}\gamma\|^{2}\\
&\mathbb{E}_{\boldsymbol{\Delta}}\left[\boldsymbol{g}_{E}^{H}\boldsymbol{\Gamma}\boldsymbol{\Gamma}^{H}\boldsymbol{g}_{E}\right]=\boldsymbol{\gamma}^{H}\left(\boldsymbol{\widehat{g}}_{E}\boldsymbol{\widehat{g}}_{E}^{H}+\sigma_{g}^{2}\boldsymbol{I}_{N}\right)\boldsymbol{\gamma}\!=\!\|\boldsymbol{R}_{E}^{1/2}\boldsymbol{\gamma}\|^{2}, 
\end{align}
wherein $\boldsymbol{R}_{E}=\boldsymbol{\widehat{g}}_{E}\boldsymbol{\widehat{g}}_{E}^{H}+\sigma_{g}^{2}\boldsymbol{I}_{N}$.
Then, \eqref{Eq:ApproxErgRate} becomes 
\begin{equation}
\mathbb{E}_{\boldsymbol{\delta}}\left[\log_{2}(1+\text{SINR}_{k,E})\right]\approx \log_{2}\left(1+\widetilde{\text{SINR}}_{k,E}\right)\;,
\end{equation}
with $\widetilde{\text{SINR}}_{k,E}$ given by 
\begin{equation}\label{Eq:SINRApprox}
\hspace{-0.2cm}\widetilde{\text{SINR}}_{k,E}\!=\!\frac{p_{k}\|\boldsymbol{R}_{E}^{1/2}\boldsymbol{H}_{k}\boldsymbol{\gamma}\|^{2}}{\sum_{m\neq k}p_{m}\|\boldsymbol{R}_{E}^{1/2}\boldsymbol{H}_{m}\boldsymbol{\gamma}\|^{2}\!+\!\sigma_{RIS}^{2}\|\boldsymbol{R}_{E}^{1/2}\boldsymbol{\gamma}\|^{2}\!+\!\sigma_{E}^{2}}
\end{equation}
Thus, the objective \eqref{Prob:aSEE_ARIS} can be approximated as 
\begin{equation}
\widetilde{\text{SEE}}=\frac{\sum_{k=1}^{K}\log_{2}(1+\text{SINR}_{k,B})-\log_{2}(1+\widetilde{\text{SINR}}_{k,E})}{P_{tot}}
\end{equation}
The next challenge is to deal with the non-convexity of the optimization problem. To this end, $\boldsymbol{\gamma}$, $\boldsymbol{p}$, and $\boldsymbol{C}_{B}$ will be optimized alternatively, as shown in the next three subsections. 

\subsection{RIS optimization}
\noindent With respect to the RIS vector $\boldsymbol{\gamma}$, the problem is expressed as:
\begin{subequations}\label{Prob:SEC_ARIS_gamma}
	\begin{align}
		&\max_{\boldsymbol{\gamma}}B\frac{\textstyle\sum_{k=1}^{K}\!\log_{2}\left(1\!+\!\text{SINR}_{k,B}\right) \!-\! \log_{2}\left(1\!+\!\widetilde{\text{SINR}}_{k,E}\right)}{\tr\left(\boldsymbol{R}\boldsymbol{\gamma}\boldsymbol{\gamma}^{H}\right)+P_{c,eq}} \label{Prob:aSEC_ARIS_gamma}\\
		&\;\text{s.t.}\;\tr\left(\boldsymbol{R}\right)\leq \tr(\boldsymbol{R}\boldsymbol{\gamma}\boldsymbol{\gamma}^{H})\leq P_{R,max}+\tr\left(\boldsymbol{R}\right)\label{Prob:bSEC_ARIS_gamma}
	\end{align}
\end{subequations}
wherein $P_{c,eq}=\sum_{k}p_{k}\mu_{k}+P_{c}-\tr(\boldsymbol{R})$. Problem \eqref{Prob:SEC_ARIS_gamma} is challenging since the objective is neither concave nor pseudo-concave, which prevents the direct use of fractional programming \cite{ZapNow15}, and \eqref{Prob:bSEC_ARIS_gamma} is a non-convex constraint. To circumvent this challenge, we employ the sequential fractional programming method. To this end, we express the term $\boldsymbol{c}_{k,B}^{H}\boldsymbol{W}_{B}\boldsymbol{c}_{k,B}$ , in terms of the vector $\boldsymbol{\gamma}$, instead of the matrix $\boldsymbol{\Gamma}$. To achieve this, we define $\boldsymbol{u}_{k}=\boldsymbol{G}_{B}^{H}\boldsymbol{c}_{k,B}$ and $\widetilde{\boldsymbol{U}}_{k,B}=\text{diag}\left(|u_{1,B}|^{2},\ldots,|u_{N,B}|^{2}\right)$. Subsequently, by incorporating the expression of $\boldsymbol{W}_{B}$, we obtain $\boldsymbol{c}_{k,B}^{H}\boldsymbol{W}_{B}\boldsymbol{c}_{k,B}=\sigma^{2}\|\boldsymbol{c}_{k,B}\|^{2}+\sigma_{\text{RIS}}^{2}\|\widetilde{\boldsymbol{U}}_{k,B}^{1/2}\boldsymbol{\gamma}\|^{2}$.
For the sequential fractional programming method, a concave lower-bound of the numerator in \eqref{Prob:aSEC_ARIS_gamma} is needed. Then, let us define 
\begin{align}
&\hspace{-0.15cm}x_{B}=p_{k}\left|\boldsymbol{c}_{k,B}^{H}\boldsymbol{A}_{k,B}\boldsymbol{\gamma}\right|^{2}\;,\; x_{E}=p_{k}\|\boldsymbol{R}_{E}^{1/2}\boldsymbol{H}_{k}\boldsymbol{\gamma}\|^{2}\\
&\hspace{-0.15cm} y_{B}\!=\!\sigma^{2}\|\boldsymbol{c}_{k,B}\|^{2}\!\!+\!\sigma_{\text{RIS}}^{2}\|\widetilde{\boldsymbol{U}}_{k,B}^{1/2}\gamma\|^{2}\!\!+\!\!\textstyle\sum_{m\neq k}\!p_{m}\!\!\left|\boldsymbol{c}_{k,B}^{H}\boldsymbol{A}_{m,B}\boldsymbol{\gamma}\right|^{2}\\
&\hspace{-0.15cm}y_{E}=\textstyle\sum_{m\neq k}p_{m}\|\boldsymbol{R}_{E}^{1/2}\boldsymbol{H}_{m}\boldsymbol{\gamma}\|^{2}+\sigma_{RIS}^{2}\|\boldsymbol{R}_{E}^{1/2}\boldsymbol{\gamma}\|^{2}\;,
\end{align}
and observe that the numerator of \eqref{Prob:aSEC_ARIS_gamma} can be written as 
\begin{align}\label{Eq:SecRate}
R_{s}&=\log_{2}\left(1+\frac{x_{B}}{y_{B}}\right)\!-\!\log_{2}\left(1\!+\!\frac{x_{E}}{y_{E}\!+\!\sigma_{E}^{2}}\right)\!=\!\log_{2}\left(1\!+\!\frac{x_{B}}{y_{B}}\right)\notag\\
&\!+\!\log_{2}\left(1\!+\!\frac{y_{E}}{\sigma_{E}^{2}}\right)\!+\!\log_{2}\left(\sigma_{E}^{2}\right)\!-\!\log_{2}(\sigma_{E}^{2}\!+\!x_{E}\!+\!y_{E})
\end{align}
Then, in order to lower-bound \eqref{Eq:SecRate}, we leverage the bounds
\begin{align}
	\log_{2}\left(1\!+\!\frac{x}{y}\right)\!\geq\! \log_{2}\left(1\!+\!\frac{\bar{x}}{\bar{y}}\right)\!+\!\frac{\bar{x}}{\bar{y}}\left(\frac{2\sqrt{x}}{\sqrt{\bar{x}}}\!-\!\frac{x+y}{\bar{x}\!+\!\bar{y}}\!-\!1\right)\\
	\log_{2}(\sigma_{E}^{2}+x+y)\leq \log(\sigma_{E}^{2}+\bar{x}+\bar{y})+\frac{x+y-\bar{x}-\bar{y}}{\sigma_{E}^{2}+\bar{x}+\bar{y}}
\end{align}
which hold for any $x$, $y$, $\bar{x}$ and $\bar{y}$, with equality whenever $x=\bar{x}$ and $y=\bar{y}$. For any feasible vector $\bar{\boldsymbol{\gamma}}$ of RIS reflection coefficients, we also define:
\begin{align}
\hspace{-0.3cm}\bar{x}_{B}&=p_{k}\left|\boldsymbol{c}_{k,B}^{H}\boldsymbol{A}_{k,B}\boldsymbol{\bar{\gamma}}\right|^{2}\;,\bar{x}_{E}=p_{k}\|\boldsymbol{R}_{E}^{1/2}\boldsymbol{H}_{k}\boldsymbol{\bar{\gamma}}\|^{2}\\
\hspace{-0.3cm}\bar{y}_{B}&\!\!=\!\!\sigma^{2}\!\|\boldsymbol{c}_{k,B}\|^{2}\!\!\!+\!\!\sigma_{\text{RIS}}^{2}\|\widetilde{\boldsymbol{U}}_{k,B}^{1/2}\boldsymbol{\bar{\gamma}}\|^{2}\!\!+\!\textstyle\sum_{m\neq k}\!p_{m}\!\left|\boldsymbol{c}_{k,B}^{H}\boldsymbol{A}_{m,B}\boldsymbol{\bar{\gamma}}\right|^{2}\\
\hspace{-0.3cm}\bar{y}_{E}&=\textstyle\sum_{m\neq k}p_{m}\|\boldsymbol{R}_{E}^{1/2}\boldsymbol{H}_{m}\boldsymbol{\bar{\gamma}}\|^{2}+\sigma_{RIS}^{2}\|\boldsymbol{R}_{E}^{1/2}\boldsymbol{\bar{\gamma}}\|^{2}
\end{align}
Then, the following lower-bound holds 
\begin{align}\label{Eq:LowerBound}
R_{s}(\boldsymbol{\gamma})&\geq \log_{2}\left(1+\frac{\bar{x}_{B}}{\bar{y}_{B}}\right)\!+\!\frac{\bar{x}_{B}}{\bar{y}_{B}}\left(\frac{2\sqrt{x_{B}}}{\sqrt{\bar{x}_{B}}}-\frac{x_{B}+y_{B}}{\bar{x}_{B}+\bar{y}_{B}}-1\right)\notag\\
&+\log_{2}\left(1+\frac{\bar{y}_{E}}{\sigma_{E}^{2}}\right)\!+\!\frac{\bar{y}_{E}}{\sigma_{E}^{2}}\left(\frac{2\sqrt{y_{E}}}{\sqrt{\bar{y}_E}}-\frac{y_{E}+\sigma_{E}^{2}}{\bar{y}_{E}+\sigma_{E}^{2}}-1\right)\notag\\
&+\log_{2}\left(\frac{\sigma_{E}^{2}}{\sigma_{E}^{2}\!+\!\bar{x}_{E}\!+\!\bar{y}_{E}}\right)-\frac{x_{E}+y_{E}-\bar{x}_{E}-\bar{y}_{E}}{\sigma_{E}^{2}+\bar{x}_{E}+\bar{y}_{E}}
\end{align}
It can be seen that all terms in \eqref{Eq:LowerBound} that depend on $\boldsymbol{\gamma}$ are concave, except for $\sqrt{x_{B}}$ and $\sqrt{y_{E}}$. However, these terms are convex in $\boldsymbol{\gamma}$, and thus they can be lower-bounded by a first-order Taylor expansion around $\boldsymbol{\bar{\gamma}}$. Let us denote by $\widetilde{R}_{s}(\boldsymbol{\gamma})$, the resulting lower-bound. 
Lastly, as for the left-hand side of \eqref{Prob:bSEC_ARIS_gamma}, since $\tr\left(\boldsymbol{R}\boldsymbol{\gamma}\boldsymbol{\gamma}^{H}\right)$ is convex in $\boldsymbol{\gamma}$, its first-order Taylor expansion around any point $\bar{\boldsymbol{\gamma}}$ provides the lower-bound $\tr\left(\boldsymbol{R}\boldsymbol{\gamma}\boldsymbol{\gamma}^{H}\right)= \boldsymbol{\gamma}\boldsymbol{R}\boldsymbol{\gamma}^{H} \geq \bar{\boldsymbol{\gamma}}\boldsymbol{R}\bar{\boldsymbol{\gamma}}^{H} + 2\Re\left\{\bar{\boldsymbol{\gamma}}^{H}\boldsymbol{R}\left(\boldsymbol{\gamma} - \bar{\boldsymbol{\gamma}}\right)\right\}$.
Consequently, each iteration of the sequential method solves:
\begin{subequations}\label{Prob:SEE_Gamma_Active}
	\begin{align}
		&\ds\max_{\boldsymbol{\gamma}}\;\frac{\textstyle\sum_{k=1}^{K}\widetilde{SR}_{k}}{\tr\left(\boldsymbol{R}\boldsymbol{\gamma}\boldsymbol{\gamma}^{H}\right)+P^{\text{(a)}}_{c,eq}} \label{Prob:aSEE_Gamma_Active}\\
		&\;\text{s.t.}\;\boldsymbol{\gamma}\boldsymbol{R}\boldsymbol{\gamma}^{H} \leq P_{R,max}+\tr\left(\boldsymbol{R}\right)\label{Prob:bSEE_Gamma_Active}\\
		&\; \quad \;\bar{\boldsymbol{\gamma}}\boldsymbol{R}\bar{\boldsymbol{\gamma}}^{H} + 2\Re\left\{\bar{\boldsymbol{\gamma}}^{H}\boldsymbol{R}\left(\boldsymbol{\gamma} - \bar{\boldsymbol{\gamma}}\right)\right\} \geq \tr\left(\boldsymbol{R}\right)\label{Prob:cSEE_Gamma_Active}
	\end{align}
\end{subequations}
The numerator and denominator of \eqref{Prob:aSEE_Gamma_Active} are concave and convex, respectively, and thus, \eqref{Prob:SEE_Gamma_Active} can be solved by fractional programming. The procedure is in Algorithm~\eqref{Alg:SCA_gamma1}, which enjoys the convergence properties of sequential programming.
\begin{algorithm}
	\caption{RIS optimization}
	\begin{algorithmic}\label{Alg:SCA_gamma1}
		\STATE $\epsilon > 0$, $\bar{\boldsymbol{\gamma}}=\boldsymbol{\gamma}_{0}$ \texttt{with} $\boldsymbol{\gamma}_{0}$ \texttt{any feasible vector};
		\REPEAT
		\STATE \texttt{Let} $\boldsymbol{\gamma}_{0}$ \texttt{be the solution of} \eqref{Prob:SEE_Gamma_Active}; \texttt{Set} $\bar{\boldsymbol{\gamma}}=\boldsymbol{\gamma}_{0}$;
		\UNTIL{$\|\bar{\boldsymbol{\gamma}}-\boldsymbol{\gamma}_{0}\|<\epsilon$}
	\end{algorithmic}
\end{algorithm}
\begin{proposition}\label{Prop:SCA_gamma1}
	Algorithm \ref{Alg:SCA_gamma1} monotonically improves the value of the objective and converges to a KKT point of \eqref{Prob:SEC_ARIS_gamma}.
\end{proposition}

\subsection{Transmit power optimization}
\noindent Let us define, for all $m$ and $k$, $a^{(B)}_{k,m} = |\boldsymbol{c}_{k,B}^{H}\boldsymbol{A}_{m,i}\boldsymbol{\gamma}|^{2}$, $d_{k,B} = \boldsymbol{c}_{k,B}^{H}\boldsymbol{W}_{B}\boldsymbol{c}_{k,B}$, $a^{(E)}_{m}=\|\boldsymbol{R}_{E}^{1/2}\boldsymbol{H}_{m}\boldsymbol{\gamma}\|^{2}$, $d_{E}=\sigma_{RIS}^{2}\|\boldsymbol{R}_{E}^{1/2}\boldsymbol{\gamma}\|^{2}+\sigma_{E}^{2}$, $P_{c,eq} = \sigma_{\text{RIS}}^{2}\left(\|\boldsymbol{\gamma}\|^{2}-N\right) + P_{c}$ and $\mu_{k,eq} = \mu_{k} - \|\boldsymbol{h}_{k}\|^{2} + \|\boldsymbol{H}_{k}\boldsymbol{\gamma}\|^{2}$. Then, the power optimization problem is formulated as in \eqref{Prob:SEE_ARIS_power}, shown at the top of this page. 
\begin{figure*}
	\begin{equation}\label{Prob:SEE_ARIS_power}
		\max_{\{p_{k}\in[0,P_{max,k}]\}_{k=1}^{K}}\,\frac{\sum_{k=1}^{K}\ds\log_{2}\left(1+\frac{p_{k}a^{(B)}_{k,k}}{d_{k,B}+\sum_{m\neq k}p_{m}a^{(B)}_{k,m}}\right)-\log_{2}\left(1+\frac{p_{k}a^{(E)}_{k}}{d_{E}+\sum_{m\neq k}p_{m}a^{(E)}_{m}}\right)}{\sum_{k=1}^{K}\mu_{k,eq}p_{k}+P_{c,eq}}
\end{equation}
\end{figure*}
Problem \eqref{Prob:SEE_ARIS_power} can be addressed again by sequential fractional programming, observing that the secrecy rate at the  numerator of \eqref{Prob:SEE_ARIS_power} can be written as the difference of two concave functions of $\boldsymbol{p}$ as 
\begin{align}
\hspace{-0.2cm}R_{s}(\boldsymbol{p})&\!\!=\!\!\underbrace{\!\!\sum_{k=1}^{K}\log\!\left(\!d_{k,B}\!+\!\!\!\sum_{m=1}^{K}p_{m}a^{(B)}_{k,m}\right)\!+\!\log\!\left(\!d_{E}\!+\!\!\!\sum_{m\neq k}p_{m}a^{(E)}_{m}\!\!\right)}_{g_{1}(\boldsymbol{p})}\notag\\
&\hspace{-1cm}\!-\!\underbrace{\!\left(\sum_{k=1}^{K}\!\!\log\!\left(\!d_{k,B}\!+\!\!\!\sum_{m\neq k}p_{m}a^{(B)}_{k,m}\right)\!\!+\!\log\!\left(\!d_{E}\!+\!\!\!\sum_{m=1}^{K}\!p_{m}a^{(E)}_{m}\right)\!\right)}_{g_{2}(\boldsymbol{p})}
\end{align}
Then a concave lower-bound of $R_{s}(\boldsymbol{p})$, can be obtained by linearizing $g_{2}(\boldsymbol{p})$ around any feasible point $\bar{\boldsymbol{p}}$, namely $R_{s}(\boldsymbol{p})\geq g_{1}(\boldsymbol{p})-g_{2}(\boldsymbol{\bar{p}})-\nabla g_{2}(\bar{\boldsymbol{p}}))^{T}(\boldsymbol{p}-\bar{\boldsymbol{p}})=\widetilde{R}_{s}(\boldsymbol{p})$
Then, a surrogate problem that fits the assumptions of sequential fractional programming can be formulated as 
	\begin{align}\label{Prob:SEE_ARIS_power_approx}
		&\max_{\{p_{k}\in[0,P_{max,k}]\}_{k=1}^{K}}\,\frac{\widetilde{R}_{s}(\boldsymbol{p})}{\sum_{k=1}^{K}\mu_{k,eq}p_{k}+P_{c,eq}}
	\end{align}
which can be solved by standard fractional programming methods. Then, the power allocation subroutine is stated as in Algorithm \ref{Alg:SCA_p1}, which is guaranteed to converge to a point fulfilling the KKT optimality conditions of \eqref{Prob:SEE_ARIS_power}. 

\begin{algorithm}
	\caption{Power optimization}
	\begin{algorithmic}\label{Alg:SCA_p1}
		\STATE $\epsilon > 0$, $\bar{\boldsymbol{p}}=\boldsymbol{p}_{0}$ \texttt{with} $\boldsymbol{p}_{0}$ \texttt{any feasible vector};
		\REPEAT
		\STATE \texttt{Let} $\boldsymbol{p}_{0}$ \texttt{be the solution of} \eqref{Prob:SEE_ARIS_power_approx}; \texttt{Set} $\bar{\boldsymbol{p}}=\boldsymbol{p}_{0}$;
		\UNTIL{$\|\bar{\boldsymbol{p}}-\boldsymbol{p}_{0}\|<\epsilon$}
	\end{algorithmic}
\end{algorithm}
\begin{proposition}\label{Prop:SCA_p1}
	Algorithm \ref{Alg:SCA_p1} monotonically improves the value of the objective and converges to a point fulfilling the KKT optimality conditions of \eqref{Prob:SEE_ARIS_power}.
\end{proposition}

\subsection{Receive filter optimization}
\noindent The optimization of $\boldsymbol{C}_{B}$, exclusively influences the legitimate rate at the numerator of the SEE, and it can be decoupled across users, boiling down to maximizing the users' individual legitimate rates. This yields the linear MMSE receiver, which, for the present case, is  $\boldsymbol{c}_{k,B}=\sqrt{p}_{k}\boldsymbol{M}_{k,B}^{-1}\boldsymbol{A}_{k,B}\boldsymbol{\gamma}$ where $\boldsymbol{M}_{k,B}=\sum_{m\neq k}p_{m}\boldsymbol{A}_{m,i}\boldsymbol{\gamma}\boldsymbol{\gamma}^{H}\boldsymbol{A}_{m,B}^{H}+\boldsymbol{W}_{B}$ represents the interference-plus-noise covariance matrix of user $k$.

\subsection[Overall Algorithm, Convergence and Complexity]{Overall  Algorithm, Convergence, and  Complexity}
The overall alternating maximization algorithm can be stated as in Algorithm~\ref{Alg:SEE1}. 
\begin{algorithm}[!h]
	\caption{Solution algorithm for Problem \eqref{Prob:SEE_ARIS}}
	\begin{algorithmic}\label{Alg:SEE1}
		\STATE \texttt{Set} $\epsilon > 0$, \texttt{initialize} $\tilde{\boldsymbol{p}}$, $\tilde{\boldsymbol{\gamma}}$ \texttt{to 
			feasible values}
		\STATE \texttt{Compute} $\boldsymbol{c}_{k}=\sqrt{p}_{k}\boldsymbol{M}_{k}^{-1}\boldsymbol{A}_{k}$ \texttt{for all} $k$;
		\REPEAT
		\STATE \texttt{Compute} $\text{SEE}_{in}=\text{SEE}(\tilde{\boldsymbol{p}},\tilde{\boldsymbol{\gamma}},\boldsymbol{C})$;
		\STATE \texttt{Given} $\tilde{\boldsymbol{p}}$ \texttt{run} Algorithm \ref{Alg:SCA_gamma1} \texttt{with output} $\tilde{\boldsymbol{\gamma}}$;
		\STATE \texttt{Given} $\tilde{\boldsymbol{\gamma}}$ \texttt{run} Algorithm \ref{Alg:SCA_p1} \texttt{with output} $\tilde{\boldsymbol{p}}$;
		\STATE \texttt{Compute} $\boldsymbol{c}_{k}\!\!=\!\!\sqrt{p}_{k}\boldsymbol{M}_{k}^{-1}\boldsymbol{A}_{k}$ and $\text{SEE}_{out}\!\!=\!\!\text{SEE}(\tilde{\boldsymbol{p}},\tilde{\boldsymbol{\gamma}},\boldsymbol{C})$;
		\UNTIL{$|\text{SEE}_{out}-\text{SEE}_{in}|<\epsilon$}
	\end{algorithmic}
\end{algorithm}

\begin{proposition}\label{Prop:SCA_Alt}
	Algorithm 3 monotonically increases the SEE value and converges in the value of the objective.
\end{proposition}

\begin{IEEEproof}
	Based on Propositions~\ref{Prop:SCA_gamma1} and~\ref{Prop:SCA_p1}, we can infer that Algorithm~\ref{Alg:SEE1} increases the SEE function in each step. Thus, since the SEE function has a finite maximizer, Algorithm~\ref{Alg:SEE1} eventually converges in the value of the objective.
\end{IEEEproof}

\textbf{Computational Complexity:}  Neglecting the complexity of computing the closed-form receive filters in $\boldsymbol{C}$, the complexity of Algorithm~\ref{Alg:SEE1} is obtained recalling that pseudo-concave maximizations with $n$ variables can be restated as concave maximization with $n+1$ variables and thus their complexity is polynomial in $n+1$~\cite{ZapNow15}. So, RIS and power optimization have complexity $\left(N+1\right)^\alpha$ and $\left(K+1\right)^\beta$,  respectively\footnote{The exponents $\alpha$ and $\beta$ are not known, but for generic convex problems they can be bounded between 1 and 4~\cite{BenTal2001ConvexOptimization}}, and thus the complexity of Algorithm~\ref{Alg:SEE1} is $\mathcal{C}_{1}=\mathcal{O}\left(I_{1}\left(I_{\gamma,1}\left(N+1\right)^\alpha + I_{p,1}\left(K+1\right)^\beta\right)\right)$ with  $I_{\gamma,1}$, $I_{p,1}$ and $I_{1}$ the number of iteration for Algorithms~\ref{Alg:SCA_gamma1},~\ref{Alg:SCA_p1}, ~\ref{Alg:SEE1} to converge.

\section{Numerical Analysis}
\begin{figure}[!h]
	\centering
	\includegraphics[width=0.4\textwidth]{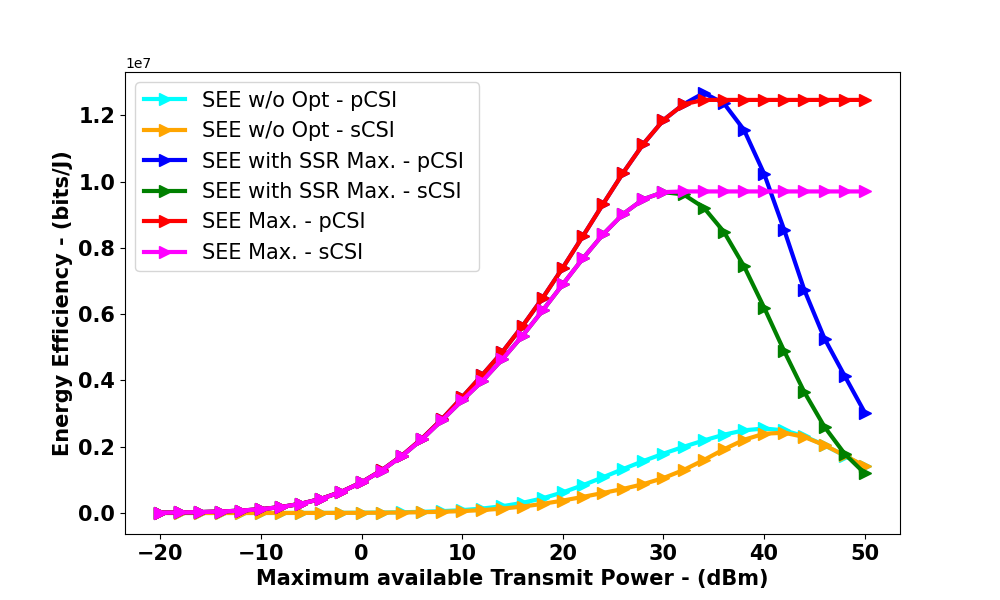}\caption{Achieved SEE versus $P_{t,max}$. $K=4$, $N_{B}=4$, $N_{E}=1$, $N=100$, $n_{h}=4, n_{g}=2$.} \label{fig:SEEvsP}
\end{figure}

\begin{figure}[!h]
	\centering
	\includegraphics[width=0.4\textwidth]{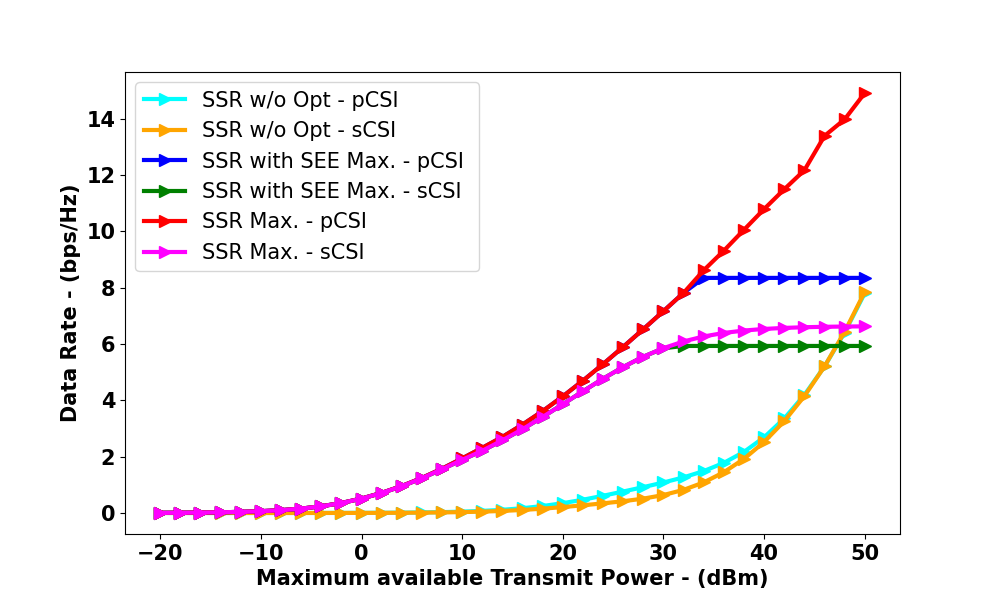}\caption{Achieved Secrecy Sum Rate versus $P_{t,max}$. $K=4$, $N_{B}=4$, $N_{E}=1$, $N=100$, $n_{h}=4, n_{g}=2$.} \label{fig:SSRvsP}
\end{figure}

For our numerical analysis, we set $K=4$, $N_{B}=4$, $N=100$, $B=20\,\textrm{MHz}$, $P_{0}=30\,\textrm{dBm}$, $P^{(a)}_{0,RIS}=20\,\textrm{dBm}$, and $P^{(p)}_{c,n}=0\,\textrm{dBm}$. The noise spectral density is  $-174\,\textrm{dBm/Hz}$ with a receive noise figure of $5\,\textrm{dB}$. Mobile users are distributed within a $30\,\textrm{m}$ radius, maintaining a minimum distance of $R_{n}=20\,\textrm{m}$ from the RIS. Bob is positioned $20\,\textrm{m}$ away from the RIS, while Eve represents any potential eavesdropping node within a $30\,\textrm{m}$ radius from Bob. Mobile users are paced at a height from $0$ to $2.5\,\textrm{m}$, whereas both the RIS and BS are $10\,\textrm{m}$ from the ground. The power decay exponents are $n_{h}=4$ and $n_{g}=2$. The channels undergo Rician fading, with factors $K_{t}=4$ for the RIS-to-BS link and $K_{r}=2$ for the user-to-RIS and RIS-to-Eve connections. The RIS-to-BS channel is allocated a higher Rician factor due to its static positioning, which bolsters the Line of Sight (LOS) component. Fig.~\ref{fig:SEEvsP} shows the achieved SEE versus $P_{t,max}$, comparing the optimal SEE with statistical CSI (SEE Max - sCSI) and perfect CSI (SEE Max - pCSI), the SEE obtained by the resource allocation which maximizes the secrecy rate with statistical CSI (SEE with SSR Max - sCSI) and perfect CSI (SEE with SSR Max - pCSI), and the SEE obtained with random power and RIS allocation with statistical CSI (w/o Opt - sCSI) and perfect CSI (w/o Opt - sCSI). It is seen that the proposed scheme largely outperforms the case in which no resource allocation is performed and that the lack of CSI does not cause detrimental SEE degradation. A similar scenario is shown in Fig.~\ref{fig:SSRvsP}, with the difference that the secrecy rate is shown. Similar remarks as for Fig.~\ref{fig:SEEvsP} hold, except that, for high $P_{max}$, not optimizing the RIS coefficients provides a higher secrecy rate than optimizing with statistical CSI.



\vspace{-0.3cm}
\section{Conclusion}
\noindent This work has proposed a provably convergent and low complexity optimization algorithm for the maximization of the SEE in the uplink of a wireless network aided by a reconfigurable intelligent surface. The analysis has shown that the lack of perfect CSI as to the eavesdropper's channel from the RIS does not cause significant SEE degradation. 


\bibliographystyle{IEEEtran}
\bibliography{FracProg.bib, references.bib}

\begin{thebibliography}{10}
\providecommand{\url}[1]{#1}
\csname url@samestyle\endcsname
\providecommand{\newblock}{\relax}
\providecommand{\bibinfo}[2]{#2}
\providecommand{\BIBentrySTDinterwordspacing}{\spaceskip=0pt\relax}
\providecommand{\BIBentryALTinterwordstretchfactor}{4}
\providecommand{\BIBentryALTinterwordspacing}{\spaceskip=\fontdimen2\font plus
\BIBentryALTinterwordstretchfactor\fontdimen3\font minus
  \fontdimen4\font\relax}
\providecommand{\BIBforeignlanguage}[2]{{%
\expandafter\ifx\csname l@#1\endcsname\relax
\typeout{** WARNING: IEEEtran.bst: No hyphenation pattern has been}%
\typeout{** loaded for the language `#1'. Using the pattern for}%
\typeout{** the default language instead.}%
\else
\language=\csname l@#1\endcsname
\fi
#2}}
\providecommand{\BIBdecl}{\relax}
\BIBdecl

\bibitem{DiRenzo2020}
M.~{Di Renzo} \emph{et~al.}, ``Analytical modeling of the path-loss for
  reconfigurable intelligent surfaces - anomalous mirror or scatterer,''
  \emph{https://arxiv.org/pdf/2001.10862.pdf}, 2020.

\bibitem{SmartWireless}
------, ``Smart radio environments empowered by reconfigurable {AI}
  meta-surfaces: An idea whose time has come,'' \emph{EURASIP Journal on
  Wireless Communincations and Networking}, vol. 129, 2019.

\bibitem{RuiZhang_COMMAG}
Q.~Wu and R.~Zhang, ``Towards smart and reconfigurable environment: Intelligent
  reflecting surface aided wireless network,'' \emph{IEEE Communications
  Magazine}, 2019.

\bibitem{Huang2019}
C.~Huang, G.~C. Alexandropoulos, C.~Yuen, and M.~Debbah, ``Indoor signal
  focusing with deep learning designed reconfigurable intelligent surfaces,''
  \emph{https://arxiv.org/pdf/1905.07726.pdf}, 2019.

\bibitem{Lopezperez2021survey}
D.~David Lopez-Perez, A.~De~Domenico, N.~Piovesan, G.~Xinli, H.~Bao, S.~Qitao,
  and M.~Debbah, ``A survey on 5g radio access network energy efficiency:
  Massive mimo, lean carrier design, sleep modes, and machine learning,''
  \emph{IEEE Communications Surveys and Tutorials}, vol.~24, no.~1, pp.
  653--697, 2022.

\bibitem{Long2021}
R.~Long, Y.-C. Liang, Y.~Pei, and E.~G. Larsson, ``Active reconfigurable
  intelligent surface-aided wireless communications,'' \emph{IEEE Transactions
  on Wireless Communications}, vol.~20, no.~8, pp. 4962--4975, 2021.

\bibitem{Fotock2023}
R.~K. Fotock, A.~Zappone, and M.~Di~Renzo, ``Energy efficiency optimization in
  ris-aided wireless networks: Active versus nearly-passive ris with global
  reflection constraints,'' \emph{IEEE Transactions on Communications}, pp.
  1--1, 2023.

\bibitem{Li2023}
X.~Li, Y.~Zheng, M.~Zeng, Y.~Liu, and O.~A. Dobre, ``Enhancing secrecy
  performance for star-ris noma networks,'' \emph{IEEE Transactions on
  Vehicular Technology}, vol.~72, no.~2, pp. 2684--2688, 2023.

\bibitem{Pei2023}
Y.~Pei, X.~Yue, W.~Yi, Y.~Liu, X.~Li, and Z.~Ding, ``Secrecy outage probability
  analysis for downlink ris-noma networks with on-off control,'' \emph{IEEE
  Transactions on Vehicular Technology}, vol.~72, no.~9, pp. 11\,772--11\,786,
  2023.

\bibitem{Zhang2022}
Z.~Zhang, J.~Chen, Y.~Liu, Q.~Wu, B.~He, and L.~Yang, ``On the secrecy design
  of star-ris assisted uplink noma networks,'' \emph{IEEE Transactions on
  Wireless Communications}, vol.~21, no.~12, pp. 11\,207--11\,221, 2022.

\bibitem{Wei2022}
L.~Wei, K.~Wang, C.~Pan, and M.~Elkashlan, ``Secrecy performance analysis of
  ris-aided communication system with randomly flying eavesdroppers,''
  \emph{IEEE Wireless Communications Letters}, vol.~11, no.~10, pp. 2240--2244,
  2022.

\bibitem{Zhao2023}
M.-M. Zhao, K.~Xu, Y.~Cai, Y.~Niu, and L.~Hanzo, ``Secrecy rate maximization of
  ris-assisted swipt systems: A two-timescale beamforming design approach,''
  \emph{IEEE Transactions on Wireless Communications}, vol.~22, no.~7, pp.
  4489--4504, 2023.

\bibitem{Hoang2023}
T.~M. Hoang, C.~Xu, A.~Vahid, H.~D. Tuan, T.~Q. Duong, and L.~Hanzo,
  ``Secrecy-rate optimization of double ris-aided space–ground networks,''
  \emph{IEEE Internet of Things Journal}, vol.~10, no.~15, pp.
  13\,221--13\,234, 2023.

\bibitem{Yizhi2023}
Y.~Li, Y.~Zou, J.~Zhu, B.~Ning, L.~Zhai, H.~Hui, Y.~Lou, and C.~Qin, ``Sum
  secrecy rate maximization for active ris-assisted uplink simo-noma
  networks,'' \emph{IEEE Communications Letters}, pp. 1--1, 2023.

\bibitem{Yichi2023}
Y.~Zhang, Y.~Lu, R.~Zhang, B.~Ai, and D.~Niyato, ``Deep reinforcement learning
  for secrecy energy efficiency maximization in ris-assisted networks,''
  \emph{IEEE Transactions on Vehicular Technology}, vol.~72, no.~9, pp.
  12\,413--12\,418, 2023.

\bibitem{Yang2023}
Y.~Lu, ``Secrecy energy efficiency in ris-assisted networks,'' \emph{IEEE
  Transactions on Vehicular Technology}, vol.~72, no.~9, pp. 12\,419--12\,424,
  2023.

\bibitem{ZapTSP13}
A.~Zappone, P.~Cao, and E.~A. Jorswieck, ``Energy efficiency optimization in
  relay-assisted {MIMO} systems with perfect and statistical {CSI},''
  \emph{IEEE Transactions on Signal Processing}, vol.~62, no.~2, pp. 443--457,
  January 2014.

\bibitem{Renzo2022}
M.~Di~Renzo, F.~H. Danufane, and S.~Tretyakov, ``Communication models for
  reconfigurable intelligent surfaces: From surface electromagnetics to
  wireless networks optimization,'' \emph{Proceedings of the IEEE}, vol. 110,
  no.~9, pp. 1164--1209, 2022.

\bibitem{Robert2023}
R.~K. Fotock, A.~Zappone, and M.~Di~Renzo, ``Energy efficiency in ris-aided
  wireless networks: Active or passive ris?'' in \emph{ICC 2023 - IEEE
  International Conference on Communications}, 2023, pp. 2704--2709.

\bibitem{ZapNow15}
A.~Zappone and E.~A. Jorswieck, ``Energy efficiency in wireless networks via
  fractional programming theory,'' \emph{Found. and Trends{\textregistered} in
  Commun. and Inf. Theory}, vol.~11, no. 3-4, pp. 185--396, 2015.

\bibitem{BenTal2001ConvexOptimization}
A.~Ben-Tal and A.~Nemirovski, \emph{Lectures on Modern Convex Optimization},
  ser. MPS/SIAM Series on Optimization.\hskip 1em plus 0.5em minus 0.4em\relax
  MPS/SIAM, 2001.

\end{thebibliography}

\end{document}